\documentclass[usegraphicx,usenatbib]{mn2e}
\usepackage{url}

\bibpunct{(}{)}{;}{a}{}{,}
\usepackage{aasmacros}
\usepackage{color}

\newcommand{\eg}{e.g.,\ }
\newcommand{\ie}{i.e.,\ }
\newcommand{\Msun}{M_{\odot}}

\newcommand{\Nifs}{$^{56}$Ni}

\def\lsim{\mathrel{\rlap{\lower 4pt \hbox{\hskip 1pt $\sim$}}\raise 1pt\hbox {$<$}}}
\def\gsim{\mathrel{\rlap{\lower 4pt \hbox{\hskip 1pt $\sim$}}\raise 1pt\hbox {$>$}}}

\title[Detectability of High-Redshift Superluminous Supernovae]
{Detectability of High-Redshift Superluminous Supernovae
with Upcoming Optical and Near-Infrared Surveys}

\author[Tanaka et al.]{
\parbox[t]{\textwidth}{
Masaomi Tanaka$^{1,2}$\thanks{E-mail: masaomi.tanaka@nao.ac.jp}, 
Takashi J. Moriya$^{2,3,4}$, 
Naoki Yoshida$^{2}$, and
Ken'ichi Nomoto$^{2}$
}
\vspace*{6pt}\\
$^1${\it National Astronomical Observatory of Japan, Mitaka, Tokyo 181-8588, Japan} \\
$^2${\it Institute for the Physics and Mathematics of the Universe, University of Tokyo, Kashiwa, Chiba 277-8583, Japan}\\
$^3${\it Department of Astronomy, Graduate School of Science, University of Tokyo, 7-3-1 Hongo, Bunkyo-ku, Tokyo 113-0033, Japan}\\
$^4${Research Center for the Early Universe, Graduate School of Science, University of Tokyo, 7-3-1 Hongo, Bunkyo-ku, Tokyo 113-0033, Japan}\\
}
\date{Accepted ---. Received ---}

\volume{000}

\pagerange{1\,-\,??} \pubyear{---}

\begin{document}
\maketitle

\label{firstpage}

\begin{abstract}
Observations of high-redshift supernovae (SNe)
open a novel opportunity to study the 
massive star population in the early Universe.
We study the detectability of superluminous SNe 
with upcoming optical and near-infrared (NIR) surveys.
Our calculations are based on the cosmic star formation
history, the SN occurence rate, the characteristic 
colour and the light curve of the SNe that are all calibrated 
by available observations.  
We show that 15-150 SNe up to $z \sim 4$ will be discovered
by the proposed Subaru/Hyper Suprime-Cam deep survey:
30 deg$^2$ survey with 24.5 AB mag depth in $z$-band for 3 months.
With its ultra-deep layer (3.5 deg$^2$ with 25.6 AB mag depth in $z$-band
for 4 months), 
the highest redshift can be extended to $z \sim 5$.
We further explore the detectability by upcoming NIR survey utilizing 
future satellites such as Euclid, WFIRST, and WISH.
The wide-field NIR surveys are very efficient to detect 
high-redshift SNe. 
With a hypothetical deep NIR survey for 
100 deg$^2$ with 26 AB mag depth at 1-4 $\mu$m,
at least $\sim$ 50 SNe will be discovered at $z>3$ in half a year.
The number of the detected SNe can place 
a strong constraint on the stellar initial mass function or its slope
especially at the high-mass end.
Superluminous SNe at high redshifts can be distinguished 
from other types of SNe by the long time-scale of their light curves 
in the observer's frame, the optical colours redder than other 
core-collapse SNe and the NIR colours redder than any other types of SNe.
\end{abstract}



\section{Introduction}

Core-collapse supernovae (SNe) are 
triggered by massive stars at the end of their lives.
Thanks to the high luminosity, 
SNe can be detected in distant galaxies.
By measuring the cosmic occurrence rate of core-collapse SNe, 
one can study the overall formation rate of the 
massive star population through cosmic time.
With upcoming or planned large-scale optical and near-infrared (NIR)
surveys, SNe could be detected to $z\sim 2$ or higher.
It is thus important to derive a realistic estimate for
the detectability in order to make an efficient survey strategy
for such observational programmes.

\citet{lien09} study the detectability of core-collapse SNe 
with the {\it Large Synoptic Survey Telescope} 
(LSST, \citealt{ivezic08,lsst09}).
They show that the wide survey with LSST (with the 
5 $\sigma$ $r$-band limiting magnitude of 25 mag)
can detect Type II SNe, core-collapse SNe with hydrogen features,
up to $z \sim 2$. Magnification by gravitational lensing
will boost the number of detectable SNe \citep{oguri10}.

Type II SNe at higher redshifts can be 
detected by an extremely deep NIR survey 
\citep{miralda-escude97,mesinger06}
with future satellites such as 
{\it James Webb Space Telescope} (JWST).
\citet{mesinger06} show that 
up to several thousands Type II SNe
can be detected at $z \sim 6$ 
if the survey depth reaches the flux limit of 3 nJy 
(30 mag in AB magnitude),
which can be achievable with 
long ($\sim 10^5$ seconds) integration with JWST.

It may be more promising to aim at detecting SNe that 
are bright in the rest frame optical to ultra-violet (UV) wavelengths.
Type IIn SNe, a type of SNe with narrow hydrogen emission lines,
are ideal targets to detect at high-redshifts \citep{cooke08}. 
Type IIn SNe are thought to be powered by the interaction of 
the SN ejecta with the circumstellar medium \citep{schlegel90},
and are bright in the rest frame optical to UV. 
\citet{cooke09} reported the discovery of 
Type IIn SNe at $z = 2.4$ and 2.0 in the archival
data of the legacy survey with Canada-France-Hawaii Telescope.

There is another type of possibly luminous
SNe worth mentioning here. Pair-instability SNe (PISNe)
are thought to be triggered by very massive stars with mass
in the range of 140-260 $\Msun$ at zero metallicity
\citep{heger02}, and are also expected to be very luminous
\citep{scannapieco05}.
PISNe might also occur even in the local Universe, 
as suggested by the recent discovery of SN 2007bi \citep{gal-yam0907bi,young10}.
However, PISNe are not very luminous at rest UV wavelengths
\citep{kasen11}, and thus they can be detected only at $z<2$ even
with LSST \citep{pan11}.
It is possible, in principle, to detect PISNe at $z>6$
with an extremely deep NIR survey with $K=$28.7 mag
(\citealt{scannapieco05}, see also recent papers by 
\citealt{pan11b,hummel11}), but the occurence rate of PISNe
(or the formation rate of extremely massive stars)
at such high redshifts is highly uncertain.
Interestingly, recent theoretical studies
and detailed simulations of the first generation of stars
suggest that the characteristic mass of the first stars
is not as large as the progenitor of a PISN
\citep{yoshida07,hosokawa11}. 
Then observing core-collapse SNe may be a more promising 
route to study the massive star population in the early Universe.

Recently, superluminous SNe
were discovered such as SN 2006gy
\citep{ofek07,smith0706gy,smith0806gy,agnoletto09,kawabata09}
and SN 2005ap \citep{quimby07}, 
whereas other superluminous SNe include
SNe 2003ma \citep{rest11}, 
2006oz \citep{leloudas12}, 
2008es \citep{miller09,gezari09}, 
2008fz \citep{drake10}, 
2008ma \citep{chatzopoulos11}, 
2010gx \citep[PTF10cwr,][]{pastorello10,quimby11}, 
SCP 06F6 \citep{barbary09}.
PTF09atu, PTF09cnd, PTF09cwl, \citep{quimby11},
PS1-10ky, PS1-10awh \citep{chomiuk11},
and CSS100217 \citep{drake11}.
See also \citet{richardson02} for other earlier candidates.
These SNe have an absolute magnitude of about $-22$ mag in optical.
Some of them are classified as Type IIn \citep[\eg][]{ofek07,smith0706gy},
or Type IIL, which is a subclass of Type II SNe 
with the linear light curve decline \citep{miller09,gezari09}, 
and others are hydrogen poor \citep[Type Ibc, \eg][]{pastorello10,quimby11}.

The nature of these superluminous SNe is poorly known.
Some of them are powered by the circumstellar interaction,
or by the shock breakout from the dense circumstellar medium
\citep{chevalier11,moriya11},
as suggested by the presence of narrow emission lines 
in superluminous Type IIn SNe \citep[\eg][]{agnoletto09}.
It is also argued that superluminous SNe could be 
powered by a large amount of \Nifs\, 
which is synthesized as a result of 
energetic core-collapse SNe \citep{umeda08,young10,moriya10}.
Other scenarios include the interaction 
between shells ejected by the pulsational pair-instability
\citep{woosley0706gy}, and SNe powered by a magnetar 
\citep{maeda07,kasen10,woosley10}.

Although the origin of the high luminosity is not yet
fully understood, the progenitors of the superluminous SNe 
are thought to be very massive.
The long time-scale of the light curve 
is suggestive of a long diffusion times of 
optical photons and hence of a large mass of ejecta
(for a given kinetic energy).
At least one progenitor of 
Type IIn SN is directly identified to be 
as luminous as luminous blue variables
with the zero-age main sequence mass of 
$M_{\rm ZAMS} > 50-80 \Msun$ \citep{gal-yam07,gal-yam09}.

Superluminous SNe may provide an important information
on the star formation in the high-redshift Universe.
Because superluminous SNe are likely to be triggered by the 
very massive end of the stellar population,
the relative occurence rate of such SNe
with respect to the overall star formation 
rate can be a sensitive probe of the slope
of the stellar initial mass function (IMF).
Theoretical studies suggest that
the shape of the IMF may be different in the early Universe, 
or could even be of top-heavy one \citep[\eg][]{larson98}.
Detecting superluminous SNe at high redshifts
will give a direct evidence for such non-standard IMF.

In this paper, we study the detectability of 
superluminous SNe at high redshifts with the upcoming 
wide-field surveys.
\citet{quimby11} noted that superluminous SNe are bright enough
that they can be detectable even at $z \sim 4$ 
with 8m-class telescopes, such as Subaru and LSST.
We derive a realistic estimate for the detectability
for a few particular set of surveys.
For the optical survey, we focus on the proposed
survey with the Subaru/Hyper Suprime-Cam (HSC, \citealt{miyazaki06}).
The method of mock observations is described in
Section \ref{sec:setup}.
We show that with a realistic survey strategy
we can detect superluminous SNe to $z\sim 4-5$ 
(Section \ref{sec:opt}).
More emphasis is made on the detectability 
by upcoming NIR survey utilizing future satellites, 
such as Euclid\footnote{\url{http://sci.esa.int/euclid}},
The Wide-Field Infrared Survey Telescope (WFIRST) 
\footnote{\url{http://wfirst.gsfc.nasa.gov}},
and Wide-field Imaging Surveyor for High-redshift
(WISH) \footnote{\url{http://www.wishmission.org/en/index.html}}
(Section \ref{sec:NIR}).
We study how the expected number of detection 
at $z>3$ is affected by the slope of the IMF at the massive end.
Details of the sample selection
are described in Section \ref{sec:selection}.
Finally, we give our concluding remarks in Section \ref{sec:conclusions}.

Throughout the paper, we assume the 
$\Omega_M = 0.3$, $\Omega_{\Lambda}=0.7$
and $H_0 = 70$ ${\rm km\ s^{-1}\ Mpc^{-1}}$ cosmology,
which is consistent with, e.g., \citet{komatsu11}.
The magnitudes are given in the AB magnitude 
when nothing is mentioned.

\begin{figure*}
  \begin{tabular}{cc}
  \includegraphics[scale=1.2]{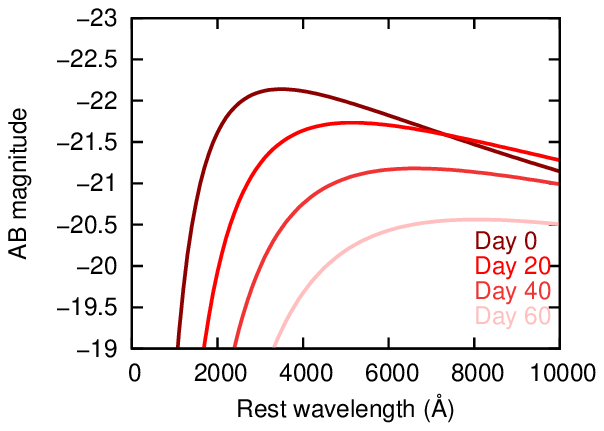} &
  \includegraphics[scale=1.2]{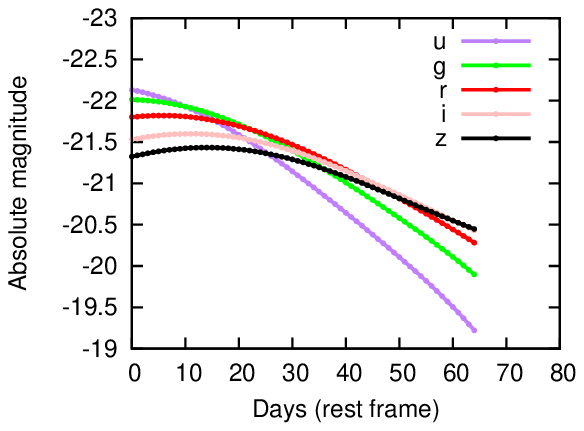} 
  \end{tabular}
  \caption{Rest-frame SED (left) and absolute light curves (right) 
of superluminous SN 2008es, used as the base model of our simulations.
The evolution of the luminosity 
and the effective blackbody temperature are taken from \citet{miller09}.}
\label{fig:SED}
\end{figure*}

\section{Setup for Mock Observations}
\label{sec:setup}

We perform simulations of mock observations 
to study the detectability of superluminous SNe at high redshift.
To this end, we need 
(1) spectral energy distributions (SEDs) of superluminous SNe
and their time evolution, 
(2) cosmic occurrence rate of superluminous SNe, and 
(3) a few key observational parameters, \ie
the depth, the area, and the cadence of observations.
In the following subsections, 
we describe the setup for our simulations.

\subsection{SED of Superluminous Supernovae}

\begin{figure}
  \includegraphics[scale=1.3]{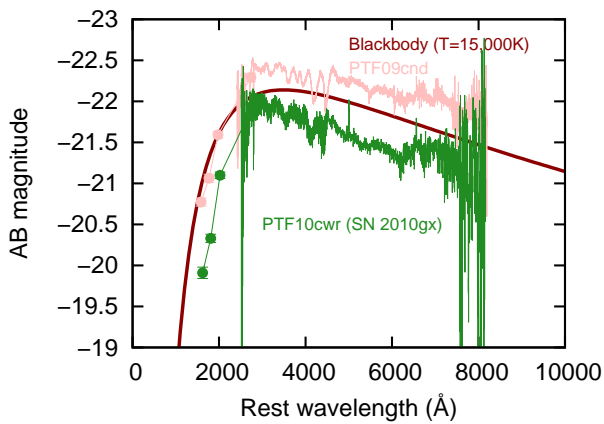} 
  \caption{SEDs of hydrogen poor superluminous SNe PTF09cnd (pink) and
PTF10cwr (SN 2010gx, green) compared with the blackbody with
$T=15,000$ K. Spectral data are from \citet{quimby11}
and photometric data are from \citet{quimby11,pastorello10}. }
\label{fig:comparison}
\end{figure}

\begin{figure*}
  \begin{tabular}{cc}
  \includegraphics[scale=1.2]{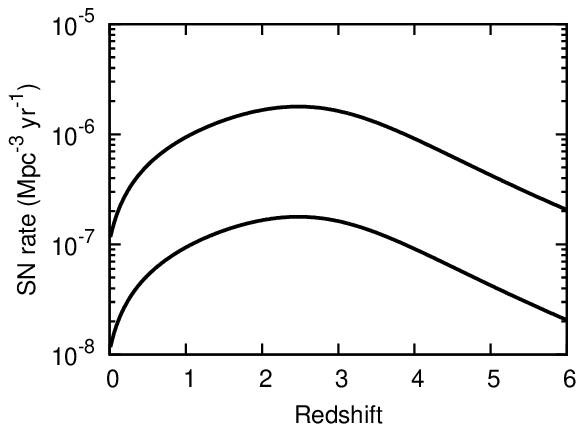} &
  \includegraphics[scale=1.2]{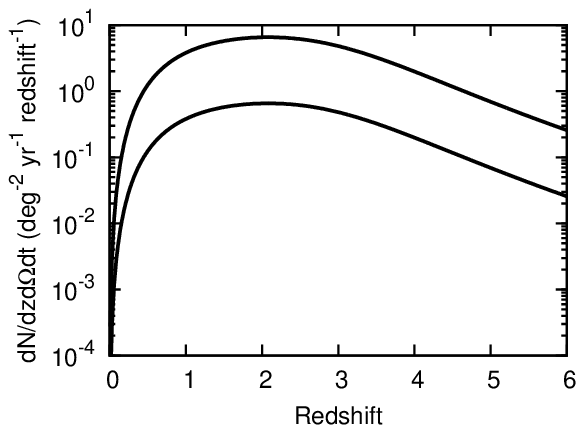} 
  \end{tabular}
  \caption{The cosmic occurrence rate of superluminous SNe
assumed in the present paper.
The left panel shows the rate per volume,
while the right panel shows the rate per area per unit redshift.
Two solid lines show the cases for $f_{\rm SLSN} = 2 \times 10^{-2}$
(top) and $2 \times 10^{-3}$ (bottom). See text for the efficiency
parameter $f_{\rm SLSN}$.}
\label{fig:SNrate}
\end{figure*}

As our fiducial model,
we use the observed SED of SN 2008es \citep{miller09,gezari09}.
SN 2008es is a well-observed object among superluminous SNe.
The SED can be approximately described by blackbody radiation with 
temperature of 6000 $-$ 15000 K depending on the phase. 
We adopt the model by \citet{miller09}
for the luminosity and the blackbody temperature.
Note that the SED is {\it not} corrected for host galaxy extinction.
We assume a cut-off of the flux at the wavelengths shorter
than the Lyman limit.
The model SED and rest-frame light curves for SN 2008es 
are shown in Figure \ref{fig:SED}.
Around the maximum brightness, the blackbody temperature is 
$\sim$ 15000 K, and the peak wavelength is 
found at about 3000 \AA.

It is known that there is a variety of SEDs of 
superluminous SNe; some SNe show H features while other do not.
Figure \ref{fig:comparison} shows 
the SEDs of hydrogen poor superluminous SNe PTF09cnd
\citep{quimby11} and PTF10cwr (SN 2010gx, \citealt{pastorello10,quimby11}).
For comparison, the blackbody spectrum with $T=15,000$ K
is also shown, which is assumed in our fiducial model at the maximum
brightness.
The overall SEDs are similar at the relevant wavelengths.
\citet{quimby11} and \citet{chomiuk11} also find that 
the evolution of the effective blackbody temperature is similar
among the superluminous SNe.
Although the SEDs of SN 2008es may not apply to
all the superluminous SNe, it is reasonable to 
assume that superluminous SNe  
have a bluer colour than other types of SNe \citep{quimby11}.

The characteristic time scale of the luminosity evolution 
differs among superluminous SNe
(see \eg Figure S1 of \citealt{quimby11}).
For example, the bolometric light curve of SN 2008es declines by 
0.85 dex in 50 days 
while that of PTF09cnd declines only 0.25 dex in the same period.
Our fiducial model is based on SN 2008es, 
which yields conservative estimates
for the number of detections.
We also perform simulations by using  
the blackbody model for PTF09cnd by \citet{quimby11}
to see the dependence on the model.

We also note that dense observations of SN 2008es were done only 
after its maximum brightness.
We do not extrapolate the light curve before the maximum
(Figure \ref{fig:SED}), and do not count SNe in the early phase.
Consequently, our simulations give conservative estimates.
We argue that, if the light curve has a symmetric shape, 
the number of detections can increase up to by a factor of 2.

SN 2008es has a peak magnitude of $-22$ mag in $g$-band,
whereas some of superluminous SNe have even a brighter magnitude 
up to $-23$ mag \citep[\eg][]{quimby07,quimby11}.
It is not yet clear if the superluminous SNe are the
objects at the luminous end of the luminosity function 
of normal core-collapse SNe, or they form an independent population.
For our simulations, we assume the latter case, 
and adopt a Gaussian luminosity function with the 
$g$-band peak magnitude of $-22$ mag.
Since the dispersion of the peak magnitude is not well constrained,
we adopt a small value (0.3 mag) so that the results are not affected
by the luminous end of the luminosity function.
Note that this luminosity function crudely includes 
the effect of extinction in the host galaxies, 
because the host extinction is {\it not} corrected in most 
of the literature.

To perform realistic mock observations, we also include 
Type Ia, Type IIP, and Type Ibc SNe in our simulations.
For the SEDs of these types, we use spectral templates by 
\citet{nugent02} 
\footnote{\url{http://supernova.lbl.gov/~nugent/nugent_templates.html}}.
For their luminosity functions, 
we assume observed average peak $R$-band Vega magnitudes 
and dispersion of nearby SNe, \ie 
$M_R = -18.6$ mag ($\sigma=$0.8 mag), $-16.1$ mag ($\sigma=$1.2 mag) 
and $-16.1$ mag ($\sigma=$1.4 mag),
for Type Ia, IIP and Ibc SNe, respectively \citep{li11}.

\subsection{Supernova Rate}

Superluminous SNe are likely to be the explosions of 
massive progenitors. 
It is reasonable to assume that the occurrence rate is 
proportional to the cosmic star formation rate $\rho_{*}(z)$
without significant time delay.
We assume the minimum zero-age main sequence mass 
for superluminous SNe to be $M_{\rm min, SLSN}=50 \Msun$, which is 
the minimum mass of luminous blue variables \citep{gal-yam07}.
Suppose only a small fraction $f_{\rm SLSN}$ of massive stars (with $>50 \Msun$) 
explode as superluminous SNe.
Then the rate of the superluminous SNe 
as a function of redshift $R_{\rm SLSN} (z)$ can be calculated as
\begin{equation}
R_{\rm SLSN}(z) 
= f_{\rm SLSN} \, \rho_{*}(z) 
\frac{ \int_{M_{\rm min, SLSN}}^{M_{\rm max, SLSN}} \psi(M) {\rm d}M }{ \int_{M_{\rm min}}^{M_{\rm max}} M \psi(M) {\rm d}M },
\label{eq:rate}
\end{equation}
where $\psi(M)$ is the stellar IMF 
\begin{equation}
\psi(M) \propto M^{-(\Gamma +1)}.
\end{equation}
We adopt a modified Salpeter A IMF of \citet{baldry03}
with the slope $\Gamma=0.5$ for $0.1\Msun < M < 0.5 \Msun$ and
$\Gamma=1.35$ for $0.5\Msun < M < 100 \Msun$.
The other parameters in Eq. (\ref{eq:rate}) are,
the minimum mass of stars $M_{\rm min}=0.1 \Msun$,
the maximum mass of stars $M_{\rm max}=100 \Msun$, 
and the maximum mass of superluminous SNe $M_{\rm max, SLSN}=100 \Msun$. 
For the star formation rate, we use the parametrized 
prescription by \citet{hopkins06}
with a modified Salpeter A IMF.

The fraction of superluminous SNe, $f_{\rm SLSN}$, is not well determined.
It is indeed one of the important goals of upcoming surveys 
to anchor the rate of such events to understand the population of progenitors.
In the present paper, we calibrate this fraction with the available 
observations so far.
\citet{quimby11} estimated the rate of hydrogen poor superluminous SNe
to be about $10^{-8}\ {\rm Mpc^{-3}\ yr^{-1}}$ at $z \simeq 0.3$, 
which is only $10^{-4}$ of total core-collapse SNe.
This fraction corresponds to $f_{\rm SLSN}=2 \times 10^{-3}$ 
of massive stars with $> 50 \Msun$.
Including hydrogen rich superluminous SNe, 
such as superluminous Type IIn SNe, the rate can be higher. 
Thus, in our simulations, 
we adopt $f_{\rm SLSN} = 2 \times 10^{-3}$ - $2 \times 10^{-2}$,
i.e., $10^{-4}$ - $10^{-3}$ of total core-collapse SNe 
are superluminous SNe.

The left panel of Figure \ref{fig:SNrate} shows 
the rate of the superluminous SNe.
The rate lies roughly between $10^{-7}$ and $10^{-6} {\rm \ Mpc^{-3}\ yr^{-1}}$
in the redshift range we are most interested in.
The SN rate per volume per year can be readily converted to 
the number per redshift per field (solid angle $\Omega$) 
and per a certain duration of observations as
\begin{equation}
\frac{{\rm d}N}{{\rm d}\Omega\, {\rm d}t \, {\rm d}z} = R_{\rm SLSN} (z) \frac{1}{1+z} \frac{{\rm d}V_{\rm com}}{{\rm d}\Omega {\rm d}z},
\end{equation}
where $V_{\rm com}$ is the comoving volume.
The right panel of Figure \ref{fig:SNrate} shows the number of 
SNe per 1 square degree, per 1 year, and per redshift.
The number of superluminous SNe at $z \sim 2$ 
is about unity within 1 deg$^2$ per year,
and does not decrease significantly to $z\sim 4$.

\begin{table*}
\begin{center}
\caption{Survey Parameters}
\label{tab:param}
\begin{tabular}{cccccccccc}
\hline
\noalign{\vspace{2pt}}
Survey                & Area              &  $\Delta t$ &  $n_1$  &    $n_2$   &  \multicolumn{5}{c}{5 $\sigma$ limiting magnitude per visit} \\
                      & (deg$^2$)        &     (day)     &         &             & $m_{g}$ & $m_{r}$ & $m_{i}$ & $m_{z}$ &  $m_{y}$\\
\noalign{\vspace{2pt}}
\hline\hline
Subaru/HSC Deep       &   30             &     6         &   2     &       3      &  26.1   & 25.8    &  25.6   & 24.5  &  23.2 \\
Subaru/HSC Ultra Deep & 3.5              &     6         &   3     &       4      &  26.9   & 26.6    &  26.6   & 25.6  &  24.3 \\
\hline
                      &                  &               &         &              & $m_{F115W}$ & $m_{F200W}$ & $m_{F277W}$ & $m_{F356W}$ & $m_{F444W}$  \\
\hline
   NIR Deep     &   100                 &     --        &   1     &         6     &  26.0      & 26.0       &  26.0      & 26.0       & 26.0      \\
\hline
\end{tabular}\\
\end{center}
\end{table*}

\subsection{Survey Strategy}

We primarily consider three types of surveys with 
different survey area, depth, cadence, 
and wavelength (Table \ref{tab:param}).
In practice, deep imaging observations must be performed 
before each survey to make a reference image.
The transient survey described below will be 
performed after such reference observations.

First, we adopt the proposed parameters for 
the deep layer of the Subaru/HSC survey (hereafter
called HSC-Deep).
The survey is planned to cover 30 deg$^2$.
The same field will be visited twice per month
with a 6-day cadence around the new-moon phase.
Such observations will be repeated for continuous 3 months.

We define the survey cadence with 3 parameters;
$n_1$ and $\Delta t$ are the number of visits and its cadence 
within one month,
and $n_2$ is the number of monthly visits.
Thus, for HSC-Deep,
$n_1 = 2$, $\Delta t = 6$ days, and $n_2 = 3$.

The HSC survey is proposed to be performed with 
5 broad band filters ($g, r, i, z$, and $y$).
The planned survey depth is summarized in Table \ref{tab:param}.
For HSC-Deep, the exposure time required to 
reach the designated depths is 
12 min for $g$ and $r$ bands, and 18 min for $i, z,$ and $y$ bands.

The second survey we consider is the Ultra Deep layer of Subaru/HSC survey
(hereafter HSC-UltraDeep).
It will observe two Subaru/HSC fields (3.5 deg$^2$ in total)
with deeper magnitude limits and more frequent cadence
(see Table \ref{tab:param}).
The required exposure times for these limits are
30 min, 30 min, 60 min, 90 min, and 90 min for $g, r, i, z,$ 
and $y$ band, respectively.
The same field will be observed 3 times per month with a 6-day cadence,
and this sequence will be repeated for 4 months, \ie
$n_1 = 3$, $\Delta t = 6$ days, and $n_2 = 4$.

Finally, we explore SN detection by NIR surveys.
We propose a hypothetical survey with a wide area of 100 deg$^2$,
which will be made possible by future satellite missions such as 
Euclid, WFIRST and WISH.
For the transmission curve of NIR filters, 
the broad band filters of JWST are used
\footnote{\url{http://www.stsci.edu/jwst/instruments/nircam/}
\url{instrumentdesign/filters/index_html}}.
We assume a constant depth of 26.0 mag for 5 bands in 1-5 $\mu$m.
For the NIR survey, we assume only one visit per month ($n_1 = 1$)
for continuous 6 months ($n_2 = 6$).

\subsection{Methods of Simulations}

 In this subsection we describe the details of our
mock observations.
We first generate superluminous SNe according to the SN rate in 
the right panel of Figure \ref{fig:SNrate} and 
the adopted survey area assumed for each survey strategy.
The observed light curve of each superluminous SN 
at a certain redshift is calculated from the SED 
in Figure \ref{fig:SED} with appropriate K-correction \citep{hogg02}.
Here the luminosity of each SN is assigned 
according to the luminosity function (with a Gaussian probability distribution).
The date of explosion is determined by a random number.
In this way, we generate expected light curves of superluminous SNe 
at a range of redshifts.

Then we perform mock observations.
The magnitude of superluminous SNe are compared with 
the limiting magnitude of the survey specified by
$\Delta t, n_1$ and $n_2$ (see Table \ref{tab:param}).
We do not consider contamination of the host galaxy, 
\ie the detection limit in Table \ref{tab:param} is kept
constant for all the objects.
Strictly speaking, this is not the case when an object is discovered 
near a bright galaxy.
However, for high-redshift superluminous SNe, the host galaxies
are expected to be faint (Section \ref{sec:selection}),
and thus it is a sound approximation to neglect the contamination.

We impose stringent detection criteria.
Only SNe that fulfill {\it both} of the following two criteria are counted
as ``detected'';
(1) brighter than 5 $\sigma$ detection limit 
in more than 2 bands at least at one epoch,
and (2) brighter than the 5 $\sigma$ detection limit 
at more than 3 epochs at least in one band.
Typically, we generate light curves for 1000 
superluminous SNe per one run.
The simulations are performed $1000$ times, 
and the number of the detection is averaged over the realizations.

\begin{figure}
\includegraphics[scale=1.2]{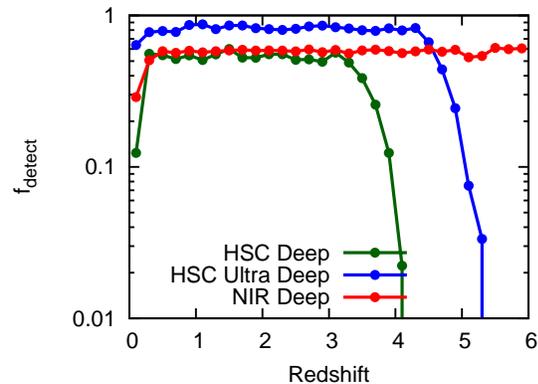} 
  \caption{Detection efficiency of superluminous SNe as a function of redshift
for HSC-Deep (green) and HSC-UltraDeep (blue), 
and for a hypothetical NIR deep survey (red).}
\label{fig:fdetect}
\end{figure}

\begin{figure}
    \includegraphics[scale=1.1]{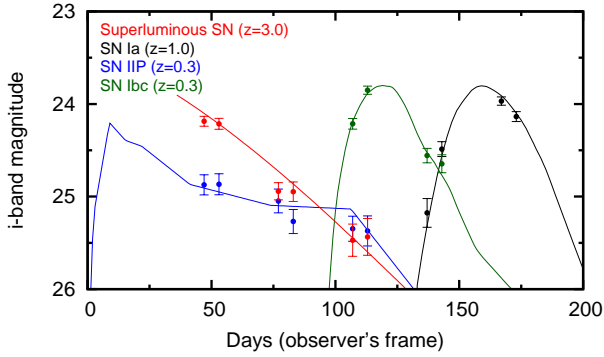} 
  \caption{An example of the simulated $i$-band light curve of 
a superluminous SN (red) for HSC-Deep.
The light curve is compared with that of Type Ia (black), 
Type IIP (blue), and Type Ibc (green) SNe at lower redshifts.
Photometric errors shown here are computed according to the limiting magnitudes
in Table \ref{tab:param}.
The light curves of Type Ia SNe and Type Ibc SNe are 
shifted toward the right for clarity.
\label{fig:LCopt}}
\end{figure}

\begin{figure}
  \begin{tabular}{c}
    \includegraphics[scale=1.2]{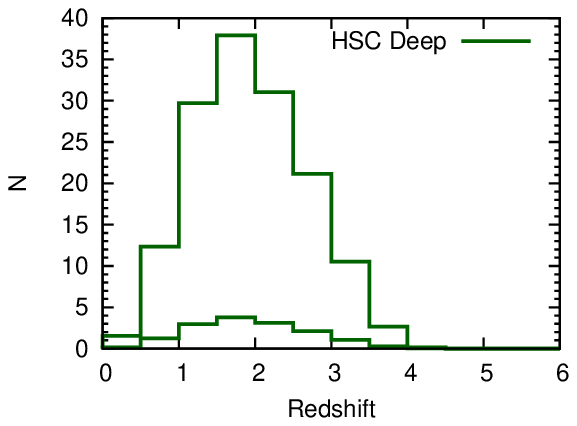} \\
    \includegraphics[scale=1.2]{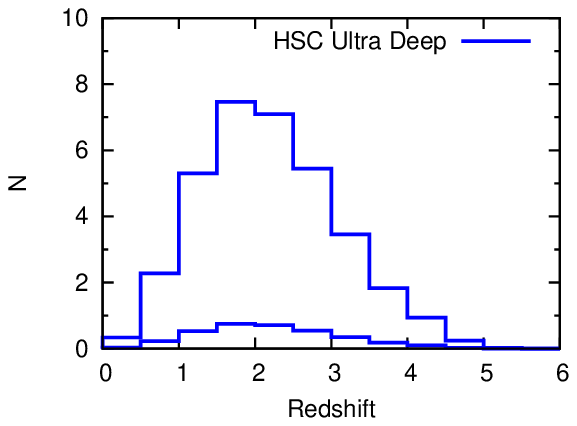}
  \end{tabular}
  \caption{
Expected number of detection as a function of redshift
for HSC-Deep (upper panel) and HSC-UltraDeep (lower panel).
The two solid lines show the cases for $f_{\rm SLSN} = 2 \times 10^{-2}$
(top) and $2 \times 10^{-3}$ (bottom).
The total number of detection is 15-150 for HSC-Deep
and 3-30 for HSC-UltraDeep.}
\label{fig:hist_opt}
\end{figure}

\section{Detectability with Optical Survey}
\label{sec:opt}

We first calculate the detection efficiency for the three surveys
we propose. 
The detection efficiency is defined to be the ratio of 
the detected number of SNe to the total number of 
SNe exploded in the survey period.
The green and blue lines in Figure \ref{fig:fdetect} show 
the detection efficiency ($f_{\rm detect}$) of superluminous SNe 
for HSC-Deep and HSC-UltraDeep, respectively.
We see rapid drop in $f_{\rm detect}$ at $z\sim 3-4$ and $z\sim 4-5$ 
for Deep and for Ultra Deep, respectively.
These limiting redshifts corresponds to the redshifts where
superluminous SNe with the mean absolute magnitude $-22$ mag becomes
undetectable in $z$-band
(see Appendix \ref{app:obsSED} for the SEDs in observer's frame).
HSC-Deep is almost complete up to $z \sim 3$ while 
HSC-UltraDeep is complete up to $z \sim 4$.
The efficiency at lower redshift is not unity
because of our somewhat stringent detection criteria.
If we impose a looser criteria, for example, 
$> 5 \sigma$ detection at only 2 epochs, 
the efficiency becomes nearly unity at low redshifts.
The efficiency and its redshift dependence are not 
largely affected even when PTF09cnd, which has a  
slower light curve than our fiducial model, 
is used as our input model.

It is worth mentioning that SNe exploded before the survey period 
can also be detected. This effect is not included in the detection 
efficiency, but such SNe are included in the detection number shown 
below.

Figure \ref{fig:LCopt} shows a few examples of
simulated light curves.
Photometric errors are computed according to the limiting magnitudes
in Table \ref{tab:param}.
The $i$-band light curve of a superluminous SNe 
is shown by the red points.
It is compared with those of other types of SNe.
Superluminous SNe at $z \sim 3$ have comparable magnitude
with other types of SNe at $z \lsim 1$.
Since the observed $i$-band wavelength corresponds 
to the rest-UV wavelengths,
the decline rate of superluminous SNe is not extremely long.
The observed time-scale is slightly longer than that of Type Ia and Ibc SNe,
but shorter than that of Type IIP SNe at plateau phase.

When PTF09cnd is used as input, the observed brightness
of a model superluminous SN at $z=3$
decreases as slowly as that of Type IIP SNe at $z=0.3$.
In this case, about 50 \% of the superluminous SNe detected with the HSC-Deep
do not show the variability larger than 1 mag in any band.
\footnote{The fraction of SNe with a small variability 
($< 1$ mag) is only $\sim$ 15 \% when the model of SN 2008es is used.
This fraction becomes smaller if the baseline the survey period is longer.
For example, if the HSC-Deep is performed over 6 months
keeping the total number of visit, \ie $n_1 = 1$ and $n_2 = 6$, 
the fraction of the small variability case is only 20 \% 
for the PTF09cnd model, and almost 0 \% for the SN 2008es model.}
It is thus important to have deep reference images 
before the survey, otherwise such SNe would be missed.

The expected number of detection of superluminous SNe 
with each layer of the survey
is shown in Figure \ref{fig:hist_opt}.
Two solid lines show the cases for $f_{\rm SLSN} = 2 \times 10^{-2}$
(top) and $2 \times 10^{-3}$ (bottom). Our reasonable guess
is that the true number will lie in between them.
The total number of detection is 
15-150 for the Deep layer, and 3-30 for the Ultra Deep layer.
If PTF09cnd is used as input, these numbers increase by 20 \%
because of the slow luminosity evolution.

For both the layers, the median redshift is found to be 
$z \sim 2$, where the star formation rate and hence
the intrinsic SN rate has a peak.
The wider survey area of HSC-Deep gives a larger 
number of detection at $z=1-3$.
In the future, a dedicated deep survey by LSST will 
discover more superluminous SNe.
If a similar depth is achieved, for example, 
in the deep drilling field \citep{lsst09},
the expected number of detection is simply proportional to the survey area.

We expect that the rate of superluminous SNe at $z \sim 1-3$
is accurately determined with HSC-Deep. 
HSC-UltraDeep can possibly detect superluminous SNe even at $z \sim 5$
if $f_{\rm SLSN} = 2 \times 10^{-2}$.
This is in contrast to PISNe, which we can detect 
only at $z<2$ with LSST \citep{pan11}.
The highest redshift of superluminous SNe depends on the intrinsic rate
of superluminous SNe, which can be observationally 
constrained by HSC-Deep.
We emphasize that surveys with the 2 layers are important 
to understand first the properties of the superluminous SNe (Deep layer)
and then to detect highest-redshift SNe (Ultra Deep layer).

\begin{figure}
    \includegraphics[scale=1.1]{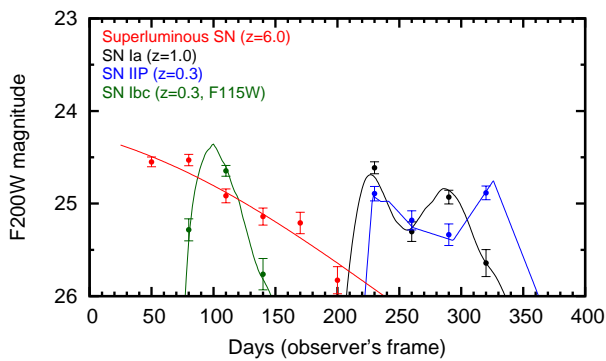} 
  \caption{
NIR light curves of superluminous SNe (red) compared with
Type Ia (black), IIP (blue), and Ibc (green) SNe.
For superluminous SNe, Type Ia and IIP SNe, magnitudes in
the F200W filter is shown.
Since the SED of Type Ibc SNe at 2$\mu$m is uncertain
(see Figure \ref{fig:spec}), magnitudes in the 
F115W filter is shown for Type Ibc SNe.
The light curves of Type Ia SNe and Type IIP SNe are 
shifted toward the right for clarity.
\label{fig:LCNIR}}
\end{figure}

\begin{figure}
  \begin{tabular}{c}
    \includegraphics[scale=1.2]{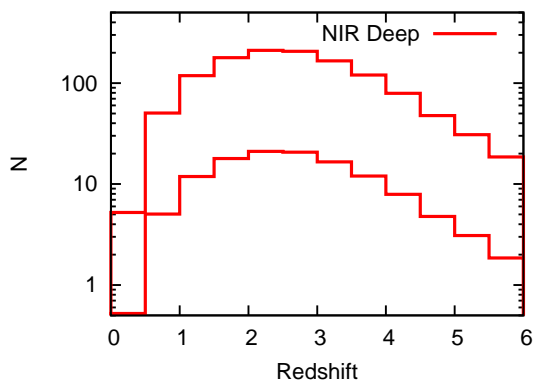}
  \end{tabular}
  \caption{Expected number of detection as a function of redshift
for the 100 deg$^2$ NIR survey.
The two solid lines show the cases for $f_{\rm SLSN} = 2 \times 10^{-2}$
(top) and $2 \times 10^{-3}$ (bottom).
The total number of detection is 120-1200.}
\label{fig:hist_NIR}
\end{figure}

\begin{figure}
  \begin{tabular}{c}
    \includegraphics[scale=1.2]{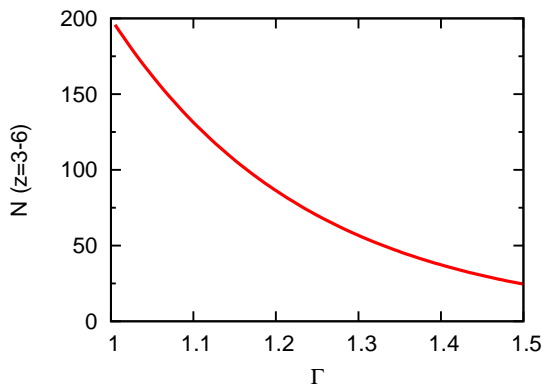}
  \end{tabular}
  \caption{Expected cumulative number of superluminous SNe 
at $z>3$ as a function of
the sloop index $\Gamma$ of the IMF at $M > 50 \Msun$.
For this plot, our 'conservative' SN rate 
($f_{\rm SLSN}=2 \times 10^{-3}$) is assumed.}
\label{fig:IMF_NIR}
\end{figure}

\section{NIR Survey}
\label{sec:NIR}
\subsection{Expected Number of Detection}

It is likely that a NIR survey is more efficient 
than the optical surveys to detect high redshift superluminous SNe.
The red line in Figure \ref{fig:fdetect} shows the detection
efficiency for our proposed NIR survey.
The efficiency is approximately constant at $z=0 - 6$.
The maximum efficiency is 0.5 simply because
we require detection at 3 epochs.
Most superluminous SNe 
up to $z \sim 6$ are above the detection limit.
However, because of our criteria, 
superluminous SNe occured in the latter 3 months 
in the survey period are not included in Figure \ref{fig:fdetect}. 
Using NIR bands is essential to perform a complete survey 
at redshift $z>3$.

Figure \ref{fig:LCNIR} shows the simulated light curves
of superluminous SNe (red) compared with other types of SNe.
Thanks to the high rest-UV luminosity (Figure \ref{fig:SED}), 
superluminous SNe even at $z=6$ can be as bright as 24-25 mag at 2 $\mu$m.
Because of the time dilation, the light curve of 
superluminous SNe at high redshifts has a long time-scale in NIR.

The expected number of detection is shown in Figure 
\ref{fig:hist_NIR}.
It is highly interesting that, with a 100 deg$^2$ survey,
hundreds of $z>2$ SNe can possibly be discovered.
The total number of SNe is 120 and 1200
for the cases of 
$f_{\rm SLSN} = 2 \times 10^{-3}$ and $2 \times 10^{-2}$, 
respectively. We discuss the usefulness of such
a large sample at high redshifts in the next subsection.

\subsection{Sensitivity to the IMF}

As we mentioned earlier, the superluminous SNe of our target 
are thought to be triggered by very massive stars. 
The occurrence rate is very sensitive to the number fraction 
of such massive stars. This in turn can be used, if the
occurence rate is accurately determined, 
to probe the slope of the IMF at the massive end 
\citep{cooke09}.
We calculate the sensitivity of the expected number 
of the detection to the slope of the stellar IMF.
For simplicity, we fix the minimum mass of 
superluminous SNe ($M_{\rm min, SLSN}$) and the fraction
of superluminous SNe ($f_{\rm SLSN}$) for the following comparison.
Note that these parameters can be observationally constrained by the 
optical survey as we discussed in Section 3.
To study the influence of the relative fraction of massive stars,
we change the slope of the IMF at the large mass $>50 \Msun$,
whereas keeping the slope at $< 50 \Msun$ to be 1.35. 
It is important to note that changing the IMF at the high-mass end 
has little impact to the calibration 
of the star formation rate since most UV continuum emission
comes from massive stars with $< 50 \Msun$.
Thus the occurence rate of superluminous SNe is 
a robust probe of the stellar IMF.

Figure \ref{fig:IMF_NIR} shows the cumulative number of detection
at $z = 3 - 6$ as a function of $\Gamma$, the slope of the IMF.
Here the fraction of superluminous SNe is conservatively assumed to be 
$f_{\rm SLSN} = 2 \times 10^{-3}$.
Even in this case, with $\Gamma < 1.1$, 
the expected number can be about three times as much as 
that with $\Gamma=1.35$.
Thus, if the star formation rate is independently measured,
and if the SN rate can be determined with an accuracy 
within a factor of 3, the very flat slope of the massive 
end of the IMF can be inferred, or rejected.

\begin{figure*}
  \begin{tabular}{cc}
    \includegraphics[scale=1.2]{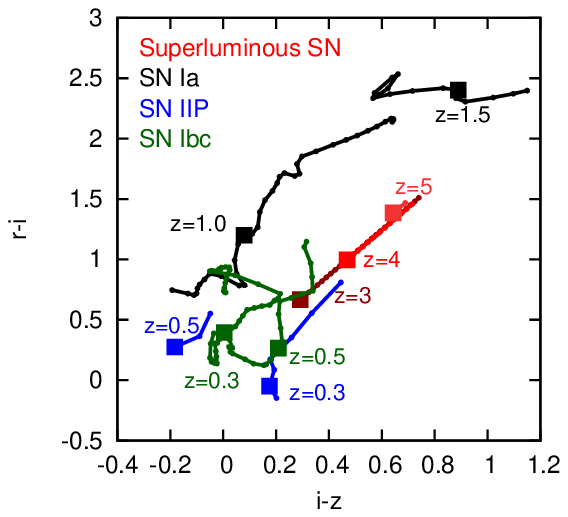} &
    \includegraphics[scale=1.2]{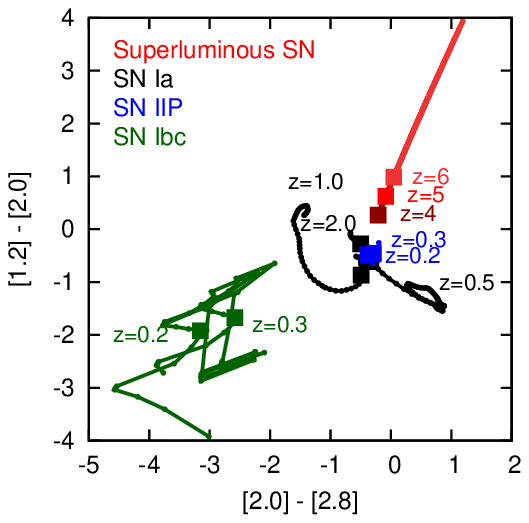} 
  \end{tabular}
  \caption{Optical (left) and NIR (right) colour-colour diagram
for various types of SNe.
Our template superluminous SN 2008es at high redshifts (red) are compared
with Type Ia SNe (black), Type IIP SNe (blue),
and Type Ibc SNe (green) at low redshifts.
The redshift of Type Ia, IIP, and Ibc SNe are selected 
so that the observed magnitudes are similar to
those of high-redshift superluminous SNe.
Colours at different epochs are connected with lines, 
and the square shows the colour at maximum brightness.
In optical, colours of superluminous SNe at high redshifts are 
red both in $r-i$ and $i-z$, 
compared with Type IIP SNe and Type Ibc SNe.
Only Type Ia SNe at redshift $z = 1.5$ can have
colours redder than superluminous SNe both in $r-i$ and $i-z$.
In NIR, superluminous SNe at high redshifts are 
redder than any other types of SNe with similar observed brightness.
For the detailed SED, see Appendix \ref{app:obsSED}.}
\label{fig:colour}
\end{figure*}

\section{Notes on the Sample Selection}
\label{sec:selection}

So far we have discussed the number of detectable SNe
of a particular type, but have not fully evaluated
the efficiency of target selection for actual surveys.
Because superluminous SNe are intrinsically rare events, 
upcoming optical surveys using 8m-class telescope, 
such as Subaru and LSST, 
and NIR surveys with Euclid, WFIRST, and WISH, 
will discover more Type Ia and ordinary core-collapse SNe.
These normal SNe will be more abundant 
than superluminous SNe by a factor of about 100.
Thus, to identify superluminous SNe at high redshifts, 
we must select the candidates properly and efficiently.

\citet{cooke08} propose a method to select
Type IIn SNe from colour-selected Lyman break galaxies.
The advantage of the method is that
the photometric redshift of the host galaxy is already known.
When Lyman break galaxies at $z>2$ are selected, 
most of SN candidates found in this sample 
are likely to be Type IIn SNe or superluminous SNe.

Here we consider the characteristics of superluminous SNe
in a more general case. An important issue
would be that the host galaxy of a distant SNe
is likely to be too faint to be detected.
We summarize the properties of (1) light curve and its time scale,
(2) colours, and (3) host galaxies of superluminous SNe at high redshifts.

The observed time-scale is determined by the 
intrinsic time-scale and the time dilation due to cosmological redshift.
Superluminous SNe observed so far indeed tend to show
an intrinsically long time-scale.
For an extreme case, 
the brightness of SN 2006gy stayed 
within 1 magnitude below the maximum for about 100 days.
Other objects such as 
2008es \citep{miller09,gezari09}, 
SCP 06F6 \citep{barbary09}.
PTF09atu, PTF09cnd, PTF09cwl \citep{quimby11}
also have a long time-scale.
Their decline tends to be faster in bluer bands (Figure \ref{fig:SED}).

In optical surveys, 
we observe high-redshift superluminous SNe at the rest-UV wavelengths.
The observed time-scale of superluminous SNe is slightly 
longer than that of Type Ia and Ibc SNe, and shorter than
Type IIP SNe (Figure \ref{fig:LCopt}),
when we use SN 2008es as a template.
When we use PTF09cnd as a template instead, 
the light curve is as slow as Type IIP SNe.
With such a slowly evolving light curve, 
it may be difficult to 
distinguish superluminous SNe from Type IIP SNe.

In NIR surveys, the long time-scale of superluminous SNe
is pronounced (Figure \ref{fig:LCNIR}).
Since NIR wavelengths traces the rest-frame optical 
at $z \gsim 2$, superluminous SNe at $z \gsim 2$ 
can stay detectable for more than 100 days
thanks to both the intrinsic brightness and the dilation effect
(Figure \ref{fig:LCNIR}).
As a result, the time-scale is longer than any other types of SNe.

The second diagnostic is the observed colour.
Figure \ref{fig:colour} shows
optical $r-i$ vs. $i-z$ (left) and 
NIR F115W-F200W ([1.2]-[2.0]) vs. F200W-F277W ([2.0]-[2.8], right)
colour-colour diagrams.
Colours of our template superluminous SN 2008es at high redshifts 
are shown in red. 
For comparison, we also show the expected colours of
Type Ia SNe at $z=1.0$ and $1.5$ (black), 
Type IIP SNe at (blue), and Type Ibc SNe (green) at $z=0.3$ and $0.5$.
These SNe are expected to have similar 
observed magnitudes to those of superluminous SNe at $z>3$.
The colours of these types of SNe are calculated by using the 
spectral templates by \citet{nugent02}.
See Appendix \ref{app:obsSED} for the SEDs.
The reddening in the host galaxy is {\it not}
taken into account here.
It is worth mentioning that there is a scatter 
in optical colour of observed SNe at low redshifts.
Including the contribution from the host galaxy, 
the scatter is typically about 1.0 mag 
for Type IIP and 0.5 mag for Type Ibc SNe \citep[\eg][]{olivares10,drout11}.
For Type Ia SNe, a wide color spread is known,
but most objects have similar colours within 0.2-0.4 mag \citep[\eg][]{wang09}.

Superluminous SNe at high redshifts tend to have 
redder colour than Type IIP and Ibc SNe both in $r-i$ and $i-z$.
For superluminous SNe at high redshifts,
the observed $r, i, z$ wavelengths correspond to 
the near-UV wavelengths in the rest frame,
which are at the blueside of the spectral peak.
On the other hand, most of Type IIP and Ibc SNe 
with optical surveys will be discovered at $z\lsim1$.
Their observed $r-i$ and $i-z$ colours still trace the 
optical wavelengths in the rest frame.
Thus, colours of superluminous SNe at high redshifts 
tend to be redder than those of 
Type IIP and Ibc SNe (see also Appendix \ref{app:obsSED}).
In optical wavelengths, only Type Ia SNe at $z \gsim 1$ have 
a similarly red or redder colour than superluminous SNe.
\footnote{It is worth mentioning here the expected optical colours of 
quasars as possible contaminations.
Most quasars at $z<4$ have $r-i=-0.5$ - $0.5$ 
and $i-z = -0.5$ - $0.5$ \citep{schneider07}, 
which are bluer than superluminous SNe at high redshifts.
Quasars at $z=4-5$ have a red $r-i$ colour (up to $\sim 2$),
which is comparable to the colour of superluminous SNe 
and Type Ia SNe.}

The red colour of superluminous SNe is more prominent 
in NIR wavelengths. See the right panel of Figure \ref{fig:colour}. 
NIR wavelengths traces the blue optical or near UV wavelength 
of superluminous SNe at high redshifts (see Appendix \ref{app:obsSED}).
For other types of SNe, observed NIR wavelengths
correspond to the red optical (Type Ia) or 
NIR (Type IIP and Ibc) in their rest frames.
Therefore, NIR colours of superluminous SNe at high redshifts
tend to be redder than those of other types of SNe.
The red colour of superluminous SNe results simply from 
the fact that they are at high redshifts.
Thus, their red color does not depend largely 
on the choice of the template SEDs.

The last characteristic is the faintness 
of the host galaxies of superluminous SNe.
An obvious reason for this is the distance to the host galaxies.
Interestingly, \citet{neill11} argue that 
the host galaxies of superluminous SNe have a faint intrinsic luminosity; 
most of them have an absolute $r$-band magnitudes of $\gsim - 20$ mag 
\footnote{Note however that there could be an observational bias that 
superluminous SNe are easily detected if their host galaxies are faint.}.
They also have quite blue UV-optical colours $NUV - r = 0$.
If there is no redshift evolution in the host galaxy population,
the optical magnitudes of the host galaxies at $z \gsim 3$
are likely to be close to the detection limit.
This is in contrast to Type Ia SNe at $z = 1-1.5$,
whose host galaxies were identified
in optical wavelengths with $m_i$ = 23-26 mag \citep{graur11}.
Overall, faintness or non-detection of the host galaxies
of the high redshift superluminous SNe may indeed
be of help with sample selection in addition to the 
characteristic features in the light curve and colour.

\section{Conclusions}
\label{sec:conclusions}

We study the detectability of superluminous SNe 
with upcoming optical and NIR surveys.
By assuming that the progenitors of superluminous SNe are 
more massive than 50 $\Msun$ and a fraction 
$f_{\rm SLSN} = 2 \times 10^{-3}$ - $2 \times 10^{-2}$
of these massive stars explode as superluminous SNe
(or the rate of superluminous SNe are $10^{-4}$ - $10^{-3}$ 
of total core-collapse SNe),
the cosmic occurrence rate of superluminous SNe 
is about 0.5-5 ${\rm deg^{-2}\ yr^{-1}\ redshift^{-1}}$
at $z\sim 1-3$ and about 0.1-1 
${\rm deg^{-2}\ yr^{-1}\ redshift^{-1}}$ at $z\sim 4-5$.

We predict that 15-150 superluminous SNe up to $z \sim 4$
will be detected with the Deep layer of the upcoming Subaru/HSC survey
if all the objects meeting our detection criteria are indeed selected.
About the half of them are at $z > 2$.
By the Deep layer, the cosmic occurrence rate of 
superluminous SNe at $z \sim 2-3$ can be anchored.
The Ultra Deep layer will discover a smaller number
of SNe (3-30), but SNe at $z\sim 5$ could be detectable.
A similarly deep, dedicated survey by LSST
will discover more superluminous SNe, 
whose number is simply proportional to the survey area.

We show that future NIR surveys will detect 
more than a few hundreds of superluminous SNe in half a year
if 100 ${\rm deg^2}$ field is surveyed 
with the depth of 26 mag.
Even with the conservative estimate of the rate
($2 \times 10^{-3}$ of massive stars with $> 50 \Msun$,
or $10^{-4}$ of total core-collapse SNe), 
such NIR survey will detect about 50 superluminous SNe at $z>3$.

Superluminous SNe at high redshifts can be selected
in the sample of high-redshift galaxies as 
demonstrated by \citet{cooke08}.
Also in the general survey, covering all redshift ranges,
superluminous SNe can be distinguished by other types of SNe 
by (1) the observed long time-scale especially, 
(2) the optical colour redder than Type IIP and Ibc SNe,
or the NIR colour redder than any other types of SNe, 
and (3) faint host galaxies.

Detection of high-redshift SNe opens
an exciting opportunity to study the 
massive star population at high redshifts.
The progenitor of superluminous SNe is likely to be 
very massive.
Therefore, the detected number is very sensitive to the IMF.
If the slope of the IMF is $\Gamma<1.1$ 
at the high mass end ($M>50 \Msun$), 
the detected number will be increased by a factor of $>$3.
If the star formation rate at $z>3$ is measured
with a small uncertainty by other methods, 
the detected number of superluminous SNe with the NIR survey
can be an indicator of top-heavy IMF at the high-redshift Universe.

\vspace{3mm}
\noindent
The authors thank Robert Quimby for giving us 
the spectra of PTF09cnd and PTF10cwr, and for fruitful discussion.
We thank the anonymous referee for valuable comments that 
improved our manuscript.
We also thank Toru Yamada, Tomoki Morokuma, Nozomu Tominaga 
and Richard Ellis for useful comments,
and Subaru/HSC transient working group for valuable discussion.
M.T., T.J.M and N.Y. are supported 
by the Japan Society for the Promotion of Science
(22840009:MT, 23.5929: TJM, 20674003:NY).
This research has been supported in part by World Premier
International Research Center Initiative, MEXT, Japan.


\appendix

\section{Observed SEDs of Supernovae}
\label{app:obsSED}

\begin{figure*}
  \begin{tabular}{cc}
    \includegraphics[scale=1.2]{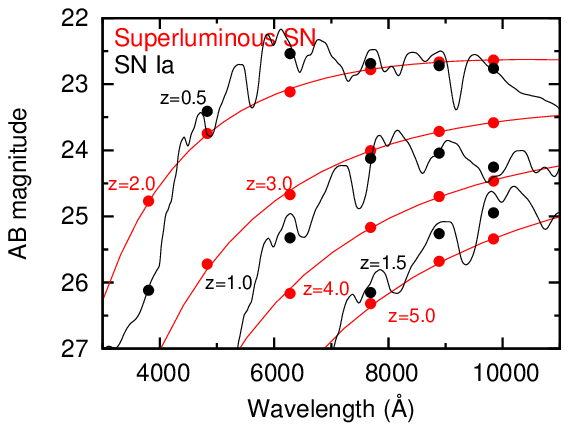} &
    \includegraphics[scale=1.2]{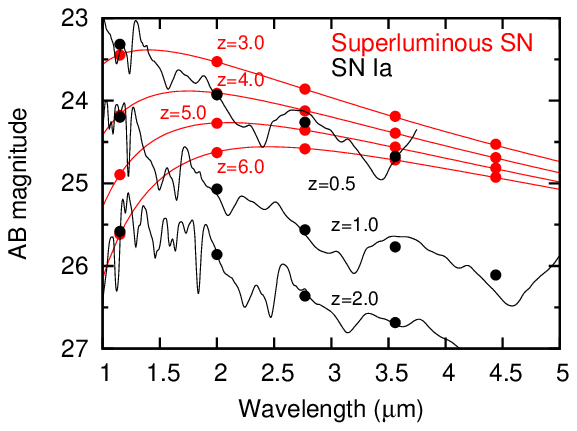} \\
    \includegraphics[scale=1.2]{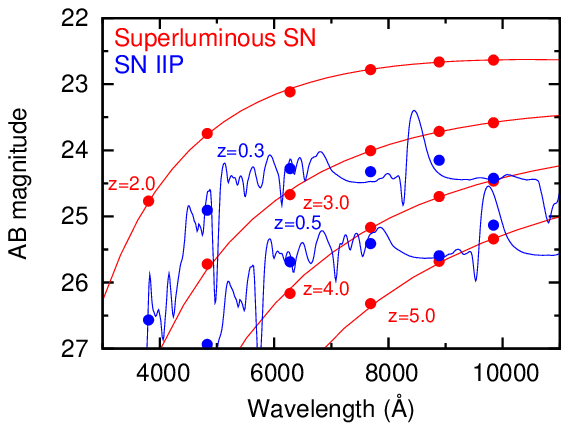} &
    \includegraphics[scale=1.2]{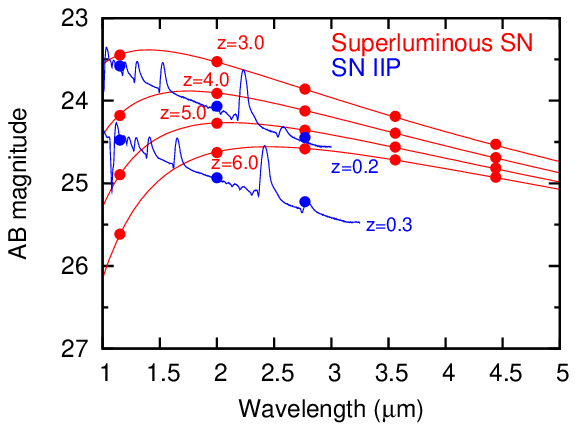} \\
    \includegraphics[scale=1.2]{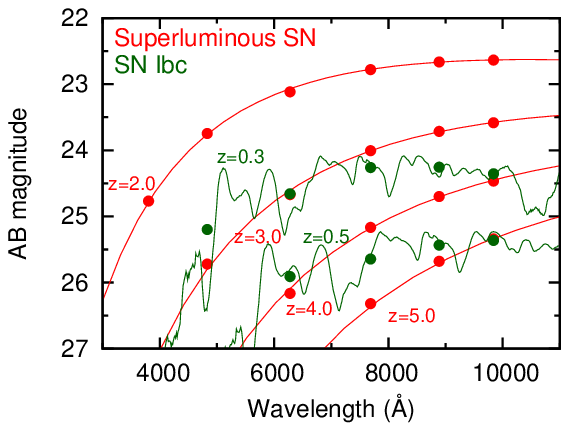} &
    \includegraphics[scale=1.2]{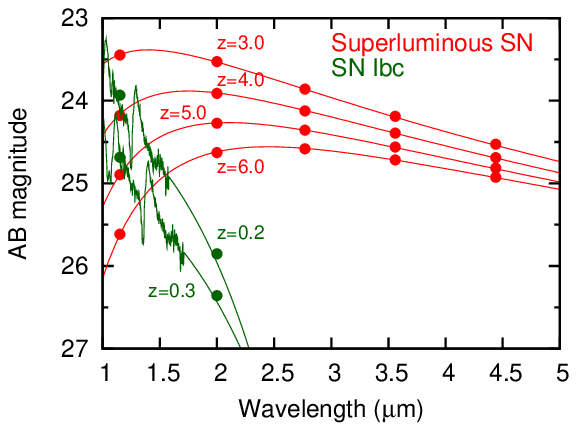}
  \end{tabular}
  \caption{SEDs of various types of SNe at maximum brightness
at different redshifts.
SEDs of superluminous SNe (red) are compared with other types of SNe
at most probable redshift ranges.}
\label{fig:spec}
\end{figure*}

Figure \ref{fig:spec} shows observed $f_{\nu}$ flux
at maximum brightness
in the unit of AB magnitude
for various types of SNe at various redshifts.
The red lines show SEDs of superluminous SN (SN 2008es).
SEDs of Type Ia SN (black), Type IIP SN (blue), and 
Type Ibc SN (green) at most probable redshifts 
are shown in comparison.
Circles shows the values of photometry
($u, g, r, i, z$, and $y$ for optical, 
and F115W, F200W, F277W, F356W, and F444W for NIR).
In this figure, 
we assume peak $V$-band Vega magnitudes that are brighter 
than the average \citep{li11}; $M_V = -19.2$ mag, 
$-16.4$ mag and $-16.4$ mag for Type Ia, IIP and Ibc SNe, 
respectively.

\end{document}